
\magnification=1250
\overfullrule=0pt
\baselineskip=19pt

\def\gs{ground state}
\def\Pa{Pauli}
\def\ma{magnetic}
\def\fd{field}
\def\chl{chiral}
\def\Ham{Hamiltonian}

\centerline{\bf Large-$U$ limit of a Hubbard model in a magnetic field:}
\centerline{\bf chiral spin interactions and paramagnetism}
\vskip.5in

\centerline{Diptiman Sen\footnote{*}{Electronic address:
diptiman@cts.iisc.ernet.in}}
\vskip .1in

\centerline{\it Centre for Theoretical Studies, Indian Institute of Science,}
\centerline{\it Bangalore 560012, India}
\vskip.2in

\centerline{R. Chitra\footnote{**}{Electronic address:
chitra@cts.iisc.ernet.in}}
\vskip .1in

\centerline{\it Department of Physics, Indian Insitute of Science,}
\centerline{\it Bangalore 560012, India}
\vskip.3in

\line{\bf Abstract \hfill}
\vskip .1in

We consider the large-$U$ limit of the one-band Hubbard model at
half-filling on a non-bipartite two-dimensional lattice. An external magnetic
field can induce a three-spin chiral interaction at order
$1 / U^2 ~$. We discuss situations
in which, at low temperatures, the chiral term may have a larger effect
than the Pauli coupling of electron spins to a magnetic field. We present a
model which explicitly demonstrates this. The ground state is a singlet
with a gap; hence the spin susceptibility is zero while the chiral
susceptibility is finite and paramagnetic.

\vskip.4in


\vfill
\eject

The \gs\ and low-energy excitations of Heisenberg antiferromagnets have
been extensively studied in
recent years. One reason for this interest is that the simplest model
for strongly correlated electron systems, the one-band Hubbard model,
reduces at half-filling to a Heisenberg antiferromagnet in
the limit of large on-site
Coulomb repulsion $U.^{1} ~$ Specifically, the Hubbard hamiltonian is
$$H ~=~- ~ \sum_{ij,\sigma} ~t_{ij} ~c_{i\sigma}^{\dag}~ c_{j\sigma}~
+ ~U ~ \sum_{i} ~n_{i\uparrow} ~n_{i\downarrow}
\eqno(1)$$
where $~t_{ij} ~= ~{t_{ji}}^{*} ~$ is the hopping integral from site
$j$ to site $i$, $\sigma ~=~ \pm ~1~ $ or, equivalently, $\uparrow$
and $\downarrow ~$, and $~n_{i\sigma} ~=~ c_{i\sigma}^{\dag} ~c_{i\sigma} ~$.
At half-filling, $~\sum_{i,\sigma} ~n_{i\sigma} ~$
is equal to the number of sites. In the large-$U$ limit,
the low-energy subspace  of the total Hilbert space  consists of states
with exactly one electron at each site. To order
$1 / U~$, this subspace is governed by the spin \Ham\
$$H_{sp} ~=~  \sum_{ij} ~J_{ij} ~{{{\bf S}_{i} \cdot
{\bf S}_{j}} \over {{\hbar}^2}}
\eqno(2)$$
Here $~{\bf S}_{i} ~=~ {({\hbar}/2) } ~c_{i\sigma}^{\dag}~
{\bf \tau}_{\sigma{ \sigma}^{\prime}} ~c_{i{\sigma}^{\prime}} ~$
with the $~{\bf \tau}~$ being Pauli spin matrices (the indices $~\sigma, ~
\sigma^{\prime} ~$ are summed over). The couplings
$~J_{ij} ~=~ 2 ~\vert ~t_{ij} ~\vert^2 ~ / U ~$ are antiferromagnetic.
At higher orders in $1 / U ~$, the $J_{ij} ~$ get renormalized
and multi-spin interactions appear in $H_{sp} ~$. These can be
calculated as an expansion in $~{t_{ij}} / U ~$ by relating Eqs.
(1) and (2) through an unitary transformation$.^{2}$

In this paper, we examine what happens when this model is placed
in a \ma\ \fd\ $.^{3} ~$ The application of the \fd\ has two effects.
Firstly, the  $~t_{ij} ~$ pick up a phase
$\exp ~[~(~{i e} / {c \hbar}~) ~\int_{i}^{j} ~{\bf A}\cdot d{\bf r} ~] ~$
where $\bf A~$ is the vector potential.
Hence the phase of the string $~ t_{ij} ~t_{jk} ~\cdot \cdot \cdot ~
t_{ki} ~$ connecting
three or more sites forming a closed curve is proportional
to the flux enclosed by that curve. Secondly, there is the \Pa\
interaction given by $~\nu ~{\bf S}_i \cdot {\bf B} / \hbar ~$
where $~\nu ~=~ - ~e \hbar /~
mc ~$, with $~e ~(m)~$ and $c$ denoting the charge (mass) of the electron
and the velocity of light respectively. (As explained below, the \Pa\
term takes the same form in the Hubbard and spin \Ham s). We are interested in
finding out whether the phases in $~t_{ij} ~$ induce any unusual
terms in $H_{sp} ~$, and what effects such terms may have. Since
an external \ma\ \fd\ breaks invariance under time-reversal $T$,
we might expect $H_{sp}~$ to reflect
this. Namely, if $~{\bf B} ~\to~ - ~{\bf B} ~$ so that  $~t_{ij} ~\to~
{t_{ij}}^{*} ~$, there should be terms in $H_{sp} ~$
which reverse sign. It turns
out that no $T$-violating terms are induced in $H_{sp}~$ on
{\it bipartite} lattices as one can show by using the
particle-hole symmetry at half-filling. We transform $~c_{i, \sigma} ~\to~
\sigma ~c_{i,~- \sigma}^{\dag} ~$ on the sites of one sublattice, and
$~ c_{i, \sigma} ~\to~ - ~\sigma ~c_{i, ~- \sigma}^{\dag}~$ on the other
sublattice. This is a symmetry of (1) if $~t_{ij} ~\to~ t_{ij}^{*} ~$
at the same time. Since $~{\bf S}_i ~$ remains invariant under this
transformation, $H_{sp}~$ must be the same for
$~\bf B~$ and $ ~- ~\bf B~$. So the $T-$ violation appears to
lie entirely in the high energy subspace (states with one or more doubly
occupied sites) for bipartite lattices. We also
observe that on {\it any} lattice, particle-hole symmetry implies that
$H_{sp}~$ must remain invariant if  $~t_{ij} ~\to~ - ~t_{ji} ~$ in (1).
Hence if the $~t_{ij} ~$ are all real, $~H_{sp}~$ cannot have odd powers of
$~t_{ij} ~$.

We must therefore consider non-bipartite lattices to obtain
something interesting. The simplest example consists  of three
sites $~i, j ~{\rm and}~ k~$ forming a triangle. The perturbative expansion in
$~t_{ij} / U ~$ is obtained by first writing the hopping term in (1) as the
sum of three terms $~T_0 ~$, $~T_1 ~$ and $~T_{-1} ~=~ {T_1}^{\dag} ~$,
where $~T_m ~$ increases the number of doubly occupied sites by $m$ when it
acts on a state$.^{2} ~$ The unitary operator relating (1) and (2) is then
given by $\exp [~i~K ~] ~$ where $K$ is a power-series in $~T_m ~/U~$.
(Note that the \Pa\ interaction commutes with
all the $~T_m ~$ and therefore with $K$).
At half-filling, the low-energy subspace (states with no doubly occupied
sites) is annihilated by both
$~T_0 ~$ and $~T_{-1} ~$ since any hopping necessarily takes us to a
state with one doubly occupied site. We then find that
$$H_{sp} ~=~ {\nu \over \hbar} ~\sum_i ~{\bf S}_i \cdot
{\bf B} ~-~ {1 \over U} ~T_{-1} ~T_1 ~+~ {1 \over U^2 } ~T_{-1} ~T_0 ~T_1
\eqno(3)$$
When rewritten in the language of spin-$1/2$ operators, the second term
on the right hand side of (3) is the same
as Eq. (2) while the third term is a three-spin \chl\ interactio$n^{3} ~$
of the form $~\mu ~{\bf S}_{i} \cdot {\bf S}_j \times {\bf S}_k /
{\hbar}^3 ~$ where
$$\mu ~=~ {{24} \over {U^2}} ~{\rm Im} ~( ~t_{ij} ~t_{jk} ~t_{ki}~)
\eqno(4)$$
This vanishes if the $~t_{ij}~$ are real. Let the \ma\ flux
enclosed by the triangle be \  $~\Phi ~$. If we denote the magnitude
$\vert ~t_{ij} ~t_{jk} ~t_{ki} ~\vert ~\equiv ~t^3 ~$ for simplicity, then
$$\mu ~=~ {{24 {t}^3} \over {U^2}} ~\sin ~( ~{{e\Phi} \over {c\hbar}} ~)
\eqno(5)$$

One can estimate the relative magnitudes of this \chl\ term and the \Pa\
term for some `typical' values of $~t_{ij}, ~U~$ and the area of the
triangle $A.^{4} ~$ For $A ~=~ 2 ~A^{o 2}$, the number $~e \Phi /
c \hbar~$ is much smaller than $1$ unless the \fd\ $B$ reaches
the fantastically large value of $~10^4 ~$ Tesla. Hence we replace the
sine in Eq. (5) by its argument, so that
$$\mu ~=~ {{24 {t}^3} \over {U^2}}~ {{eA} \over {c \hbar}}~ B ~\cos \theta
\eqno(6)$$
where $\theta$ is the angle between $~\bf B~$ and the normal
to the plane of the triangle.
We then find that for $~t ~=~ 0.5~$ eV and $~U ~=~5~$ eV, the magnitude of the
\Pa\ term is about forty times larger than the \chl\ term. This estimate
follows from comparing the splitting in the \gs\ energy produced by the
\Pa\ and \chl\ interactions for a triangle in which
the three $~J_{ij}$'s are equal.
Namely, we take the \Ham\ on a triangle to be
$$\eqalign{H_{sp} ~=~ &J~ [~{\bf S}_i \cdot {\bf S}_j ~+ {\bf S}_j \cdot
{\bf S}_k ~+~ {\bf S}_k \cdot {\bf S}_i ~] \cr
&+~ {\mu \over {\hbar}^3} ~~{\bf S}_i \cdot {\bf S}_j \times
{\bf S}_k ~+~ {\nu \over \hbar} ~(~ {\bf S}_i ~+ ~{\bf S}_j ~+~
{\bf S}_k ~) \cdot {\bf B} \cr}
\eqno(7)$$
If $~{\bf B} ~=~ 0~$, the ground state of this \Ham\ has a four-fold
degeneracy with
all states having total $~S ~=~ 1/2 ~$. The \ma\ \fd\ breaks this degeneracy
completely with the \Pa\ and \chl\ terms contributing $~\pm ~\nu B/2 ~$ and
$~\pm ~{\sqrt 3} ~\mu /4 ~$ respectively. For the excited states with $~S ~=~
3/2 ~$, we observe that the \chl\ term has no effect. Thus the \chl\
interaction can only lower the energy of a state if it is non-ferromagnetic.

Although the \Pa\ term appears to be numerically much larger than the
\chl\ term, one can think of two possible situations in which the \chl\
interaction dominates. The first example is one in which the \gs\ is a spin
singlet and is \chl\ {\it even} in the absence of the \ma\ \fd\ .
We have in mind here the two-dimensional models discussed
by Wen, Wilczek and Ze$e^{5} ~$ where a spin \Ham\ has two
degenerate singlet \gs s with opposite chiralities. We can say that
each \gs\ has a non-zero {\it chiral moment} $~M_c ~$ (defined later). Then
an applied \ma\ \fd\ picks out one of the two \gs s due to an
interaction of the form $~- ~M_c ~B ~$. The \Pa\ term
$~{\bf S} \cdot {\bf B} ~$ (where $ ~{\bf S} ~=~ \sum_i ~{\bf S}_i ~$) plays
no role here because the \gs s are singlets. However, these kinds of models
often require a special choice of the two-spin couplings $~J_{ij} ~$ as well
as peculiar multi-spin interactions in order to produce the required \gs\
degeneracy. There are also papers which argue that frustrated antiferromagnets
with only two-spin interactions can have  chiral ground states$.^{5-7}~$
However, this has been questione$d^{8} ~$
and it seems to be quite difficult to have chiral ground states in the
absence of an external \ma\ \fd\ .

The second example, which only has short-range two-spin interactions
and does not require a fine-tuning of the couplings, is one
in which the \gs\ is a singlet, unique and
{\it non-chiral} in the absence of the \ma\ \fd\ .
Further, there is a gap $\Delta$ to states with total spin greater than zero.
Then the \gs\ continues to be a singlet in the presence of a field
if $~\vert ~\nu B ~\vert ~\ll ~\Delta $.
But it may develop a \chl\ moment $~M_c ~$ to
first-order in $~B~$ and one can define a {\it chiral susceptibility}
$~\chi_c ~\equiv~
(~ {\partial M_c} / {\partial B} ~)_{B~=~0} ~$. So one may have a
finite $~\chi_c ~$ even though the {\it spin susceptibility} $~\chi_s ~=~
{\partial ~\langle~ S ~\rangle~ } / {\partial B} ~=~0 ~$.
We now present a two-dimensional model which explicitly demonstrates all this.
Incidentally, it is the only two-dimensional spin model that we are
aware of in which the ground state and the low-lying excitations can be
found exactly (for the Hamiltonian $~H_0 ~$ given below).

Our model, shown in Fig. 1, consists of chains of rhombuses which are
coupled to each other in the form of a brick lattice. Each
rhombus is formed out of two triangles with a common base. The number
of rhombuses is $ ~N/ 3~$ if the number of sites is $~N~$.
Starting from a Hubbard model with only nearest-neighbor hoppings,
the spin \Ham\ in a \ma\ \fd\ is given up to order $~1 / U^2 ~$ by
$$\eqalign{H_{sp}~ &= ~H_0 ~+ ~\mu ~{C \over {{\hbar}^3}} \cr
{\rm where} \quad H_0 ~ &= ~\sum_{ij} ~J_{ij} ~{{{\bf S}_i
\cdot {\bf S}_j} \over {{\hbar}^2}}~
+~ {\nu \over \hbar} ~\sum_{i} ~{\bf S}_i \cdot {\bf B} \cr
{\rm and} \quad ~C ~ &= ~\sum_{\alpha} ~{\bf S}_3 \cdot {\bf S}_4 \times (~
{\bf S}_2 ~- ~{\bf S}_5 ~) \cr}
\eqno(8)$$
The \chl\ term $C$ is a sum over rhombuses labelled by the index $~\alpha ~$,
with each rhombus contributing the sum of two three-spin terms as
indicated in Eq. (8). (See Fig. 1 for the site labels $1$ to $6$ in and
around a typical rhombus). Note that the \Pa\ term has been included
in an `unperturbed' \Ham\ $H_0~$, while the \chl\ term will be considered
perturbatively in the following. If we define {\it parity}
to be the transformation which exchanges
the top and bottom sites of the vertical bonds inside all the rhombuses
simultaneously (namely, $~3 ~\leftrightarrow~ 4~$ in the figure), then
$C$ is odd under parity while $H_0 ~$ is even under parity.

In $H_{sp} ~$, the couplings $~J_{ij} ~$ on the vertical bonds inside the
rhombuses, the slanted bonds, and the vertical bonds joining the chains
are denoted by $~J_1 ~, ~J_2~$
and $~J_3~$ respectively as shown in Fig. 1. Let us assume that
$~J_1 ~>~ 2 J_2~$ and that $~\vert ~\nu B ~\vert ~$ is much less than both
$(~J_1 ~- ~2 J_2 ~) ~\hbar~$ and $ ~J_3 ~{\hbar} ~$. Then one can prove that
the \gs\ of $H_0~$ is a singlet, unique and has a gap $~\Delta_0 ~$ to
all excitations. Let us first introduce the notation $~O_{ij} ~$ and
$ ~1_{ij} ~$ for the singlet
and triplet states respectively formed from the spins at sites $i$ and $j$.
Note that $~O_{ij} ~=~ {1 / {\sqrt 2}} ~( ~\vert ~i \uparrow ~j
\downarrow ~\rangle ~-~
\vert ~i \downarrow ~j \uparrow ~\rangle ~) ~$ is antisymmetric
under an exchange of $i$ and $j$, while the three states collectively denoted
by $~1_{ij}~$ are all symmetric. Then the \gs\ of $H_0~$ is the state
$~\psi_{0} ~$ given by the product of singlets
$~\cdot \cdot \cdot ~O_{12} ~\otimes ~O_{34}~ \otimes~ O_{56}
\cdot \cdot \cdot ~$ following the labels in Fig. 1. Namely,
each of the vertical bonds form a singlet. The \gs\ energy is
$E_0 ~=~ - ~(~N / 8~)~ ( ~2~ J_1 ~+ ~J_3 ~)~$.

To prove that $~\psi_0 ~$ is the \gs\ and that there is a gap $~\Delta_0 ~$,
let us write  $~H_0 ~=~ H_1 ~+ ~H_2 ~$ where $H_1 ~$ is the same as $H_0 ~$
except that the couplings $J_1 ~$ are replaced by $J_1 ~- ~2 J_2 ~$
and the couplings $J_2~$ are replaced by zero. Thus $H_1 ~$ is a sum
of disconnected two-spin \Ham s involving only the vertical bonds, while
$H_2 ~$ is a sum of disconnected four-spin \Ham s of the form
$~(J_2 ~/ 2) ~[~(~{\bf S}_2~ + ~{\bf S}_3 ~+~ {\bf S}_4 ~)^{2}~
+ ~(~{\bf S}_3 ~+ ~{\bf S}_4 ~+~ {\bf S}_5 ~)^{2} ~]~$ for each rhombus.
It is then easy to find the complete spectra for both
$H_1 ~$ and $H_2 ~$. For $H_1 ~$, $~\psi _0 ~$ is the unique
\gs\ and there is a finite gap $\Delta_1 ~=~ {\rm min} ~(~J_1 ~-~ 2 J_2 ~$ $~
- ~\vert ~\nu B ~\vert~, ~J_3 ~-~\vert ~\nu B ~\vert ~)$
to the space of states orthogonal to
$~\psi_0 ~$. For $H_2 ~$, $~\psi_0 ~$ is a \gs\ but there is no gap to its
orthogonal subspace. It then follows that $~\psi_0 ~$  is the \gs\
of $~H_1 ~+ ~H_2 ~$ and there is a gap $\Delta_0 ~\ge ~\Delta_1 ~$
to all other states$.^{9} ~$ We also note
that the total spin on any bond of type $~J_1 ~$ (e.g. $~(~{\bf S}_3 ~+~
{\bf S}_4 ~)^2 ~$) commutes with $H_0~$, so that there are $~N / 3~$
operators which can be diagonalized along with $H_0 ~$. This important
property proves to be very useful. For instance, it implies that a low-lying
excitation can only have a {\it finite} number $~p~$ of $~J_1~$ bonds forming
triplets. (Such a state is separated from the ground state by a gap $~\Delta ~
\ge ~p ~(~J_1 ~-~ 2 J_2 ~- ~\vert ~\nu B ~\vert ~)~$ by a similar argument
involving $~H_1 ~$ and $~H_2 ~$). Further, such an excitation can only differ
from the ground state in a local neighborhood of those
$~p~$ triplet bonds or due to some isolated $~J_3 ~$ bonds forming triplets
instead of singlets. Hence, all low-lying excitations are localized and
dispersionless (with energy independent of the momentum).

One can check that $~\psi_0 ~$ is non-chiral, namely,
$\langle ~\psi_{0} ~\vert ~C~ \vert ~\psi_{0} ~\rangle ~=~0 ~$. The simplest
way to show this is to write $~C ~=~ \sum_{\alpha} ~C_{\alpha} ~$, where
$~C_{\alpha} ~$ is the sum of the two \chl\ terms in a rhombus
$~\alpha$. Consider the rhombus made out of sites $2$, $3$, $4$ and $5$ in
Fig. 1. Both
$~\psi_0 ~$ and $~C_{\alpha} ~$ for that rhombus are odd under the parity
transformation $~3 ~\leftrightarrow ~4~$. Hence $~C_{\alpha} ~\vert ~\psi_{0}~
\rangle ~$ is even under parity, i.e., it contains $~1_{34} ~$ rather than
$~O_{34} ~$. Hence $~\langle ~\psi_{0} ~\vert ~C_{\alpha} ~\vert ~\psi_{0}~
\rangle ~ $ must be zero. We will now assume that the term $~\mu C ~$ in
(8) only changes the
\gs\ perturbatively because we expect the gap to survive for a finite
range of $\mu$ around $~\mu ~=~0 ~$. We can therefore use second-order
perturbation theory to compute the \gs\ energy $~E_{0}(\mu) ~$
to order $\mu^2 ~$. Since
$\mu ~$ is proportional to $B$ , this will give us the \chl\ moment
$~M_c ~ \equiv~ -~ {\partial E_{0}(\mu)} / {\partial B}$ and the
\chl\ susceptibility $~\chi_c ~=~ {\partial M_c } / {\partial B} ~$.

The second-order expression is
$$E_{0}(\mu) ~- ~E_{0} ~=~ {\mu}^{2}~
\sum_{n\ne 0} ~\sum_{\alpha, \beta} ~{{ (C_{\alpha} ~)_{0n} ~
(C_{\beta} ~)_{n0}} \over {E_{n} ~- ~E_{0}}}
\eqno(9)$$
where $~(C_{\alpha} ~)_{mn} ~=~ \langle ~\psi_{m} ~\vert ~C_{\alpha} ~\vert~
\psi_{n} ~\rangle~$ and $~m, n~$ label the eigenstates of $H_0~$. Now,
we know that the state $~C_{\beta} ~\vert ~\psi_{0} ~\rangle~$
has the $~J_1 ~$ bond in rhombus $\beta$ forming a spin triplet
while the $~J_1 ~$ bonds in all other rhombuses are singlets. This
implies that all the terms in (9) with $~\alpha ~\ne~ \beta ~$ must
vanish. Next let us consider a particular rhombus labelled $\beta ~$
and the six sites in and around that rhombus as labelled in Fig. 1.
Then $~(C_{\beta} ~)_{n0} ~$ can only be non-zero for a {\it finite} number of
states $~\psi_n ~$ since the singlet subspace
of those six spins is five-dimensional. Also, $~( C_{\beta} ~)_{n0} ~$
can be non-zero only if the vertical bond in the rhombus $\beta ~$
is a triplet in the state $~\psi_n ~$.
An explicit calculation shows that $~(C_{\beta} ~)_{n0} ~$
is actually non-zero for only two states for which $~E_n ~- ~E_0 ~$ are
given by
$$E_{\pm} ~- ~E_0 ~=~ J_1 ~- ~J_2 + ~{3 \over 2} ~J_3 ~\pm ~
{\sqrt {~ {J_2}^2 ~ +~ {{J_3}^2 \over 4}} }
\eqno(10)$$
respectively. ($~E_{\pm } ~$ and $~E_0 ~$ are independent of the magnetic
field $B$ since all the states being considered are spin singlets.)
We eventually find that
$$E_{0}(\mu) ~- ~E_0 ~=~ - ~{N \over 8} ~{\mu}^{2} ~{{J_1 ~- ~
J_2 ~+ ~2 J_3} \over {( ~E_+ ~- ~E_0 ~)~ (~ E_- ~-~ E_0 ~)}}
\eqno(11)$$
This is of order $~t^4 / U^3 ~$ since the couplings $~J_{ij} ~
\sim ~{t^2} / U ~$.
This and Eq. (6) yield a {\it paramagnetic} susceptibility
$$\chi_c ~=~ {N \over 4} ~ \biggl( ~{{24 {t}^{3}} \over {U^2}}~ {{eA}
\over {c \hbar}} ~\cos \theta \biggr)^2 ~~{{J_1 ~- ~J_2 ~+ ~2J_3 }
\over {( ~E_+ ~-~ E_0 ~) ~( ~E_- ~ - ~E_0 ~)}}
\eqno(12)$$
to order $ ~t^4 / U^3 ~$.

We emphasize that (11) is not the complete expression for the \gs\
energy to order $~1 / {U^3} ~$, since we have ignored terms
of order $1 / {U^3} ~$ in deriving  $H_{sp} ~$ in (8). These terms do
contribute to the energy in {\it first-order} perturbation theory. However,
one can show that these terms are not \chl\ because they are
even under parity. They are independent of $\mu$ and hence
do not contribute to the \chl\ quantities $M_c ~$ and $\chi_c ~$
to order $~ t^4 / U^3 ~$.

At finite temperatures, this model no longer has $ ~\chi_s ~=~0 ~$. But if the
temperature is small compared to the gap, then $~\chi_c ~$ will continue to\
be much larger than $~\chi_s ~$.

Our model is somewhat peculiar in that the two-spin correlation
in the \gs\ of $H_0 ~$ is {\it exactly} zero beyond a
short distance. However, this
property is unlikely to survive once  we take into account
the \chl\ terms and higher order terms in $~1 / U ~$ which couple spins
on non-neighboring sites. We then expect the two-spin correlation
to go to zero exponentially at large separations because of the gap
above the \gs\ . The \gs\ is therefore a spin-liquid which is
dominated by short-range valence bonds.

More realistic models which do not have a gap to spin excitations
will generally have both $~\chi_c ~$ and $~\chi_s ~$ non-zero even at
zero temperature. For instance, one can consider a Hubbard model on a
triangular lattice, or on a square
lattice with both nearest-neighbor and next-nearest-neighbor hoppings.
Whether $~\chi_c ~$ will be comparable to or much
smaller than $~\chi_s ~$ will then depend on the properties of low-energy
excitations in the absence of the \ma\ \fd\ . For instance, if there
are singlet \chl\ states lying very close to a
non-chiral \gs , then one
would expect $~\chi_c ~$ to be large. An important (and perhaps
experimentally observable)  difference between the two susceptibilites
is that $~\chi_c ~$ depends on the orientation of the magnetic field
with respect to the plane containing the sites of the spins.

To conclude, we have seen that a spin system which arises from an
underlying Hubbard model can develop \chl\ interactions when placed in
a \ma\ \fd .
Although these interactions are small, they may lead to an interesting
low-temperature phase resembling a \chl\ spin-liquid.

\vskip .4in

\line{\bf Acknowledgments \hfill}
\vskip .2in

We thank D. M. Gaitonde, H. R. Krishnamurthy and B. S. Shastry for useful
discussions.

\vskip .4in

\line{\bf References \hfill}
\vskip .2in

\noindent
\item{1.}{P. W. Anderson, Science {\bf 235}, 1196 (1987) and Phys.
Rev. {\bf 115}, 2 (1959).}

\noindent
\item{2.}{A. H. MacDonald, S. M. Girvin and D. Yoshioka, Phys. Rev. B {\bf 37},
9753 (1988); C. Gros, R. Joynt and T. M. Rice, Phys. Rev. B
{\bf 36}, 381 (1987); J. E. Hirsch, Phys. Rev. Lett. {\bf 54}, 1317 (1985);
M. Takahashi, J. Phys. C {\bf 10}, 1289 (1977).}

\noindent
\item{3.}{D. S. Rokhsar, Phys. Rev. Lett. {\bf 65}, 1506 (1990); J. K.
Freericks, L. M. Falicov and D. S. Rokhsar, Phys. Rev. B {\bf 44}, 1458
(1991).}

\noindent
\item{4.}{P. W. Anderson, G. Baskaran, Z. Zou and T. Hsu, Phys. Rev. Lett
{\bf 58}, 2790 (1987).}

\noindent
\item{5.}{X. G. Wen, F. Wilczek and A. Zee, Phys. Rev. B {\bf 39}, 11413
(1989).}

\noindent
\item{6.}{V. Kalmeyer and R.B. Laughlin, Phys. Rev. Lett.
{\bf 59}, 2095 (1987).}

\noindent
\item{7.}{G. Baskaran, Phys. Rev. Lett. {\bf 63}, 2524 (1989); I. Ritchey,
P. Chandra and P. Coleman, {\it ibid.} {\bf 64}, 2583 (1990);
G. Baskaran, {\it ibid.} {\bf 64}, 2584 (1990).}

\noindent
\item{8.}{E. T. Tomboulis, Phys. Rev. Lett. {\bf 68}, 3100 (1992); S. Yu.
Khlebnikov, JETP Lett. {\bf 53}, 88 (1991); E. Dagotto and A. Moreo, Phys.
Rev. Lett. {\bf 63}, 2148 (1989).}

\noindent
\item{9.}{P. W. Anderson, Phys. Rev. {\bf 83}, 1260 (1951).}

\vfill
\eject

\line{\bf Figure Caption \hfill}
\vskip .2in

\noindent
\item{1.}{The model showing six sites labelled $1$ to $6$ in and around
a typical rhombus. The three different antiferromagnetic couplings $~J_1 ~$,
$~J_2 ~$ and $~J_3 ~$ are also indicated.}

\end